%% file: main.tex
\pdfoutput=1
\documentclass[11pt]{article}

\usepackage[preprint]{acl}

\usepackage{adjustbox}
\usepackage{times}
\usepackage{latexsym}
\usepackage{mathtools}
\usepackage{xspace}
\usepackage{caption}
\usepackage{subcaption}
\usepackage[utf8]{inputenc} % allow utf-8 input
\usepackage[T1]{fontenc}    % use 8-bit T1 fonts
\usepackage{hyperref}       % hyperlinks
\usepackage{url}            % simple URL typesetting
\usepackage{booktabs}       % professional-quality tables
\usepackage{nicefrac}       % compact symbols for 1/2, etc.
\usepackage{microtype}      % microtypography
\usepackage{lipsum}
\usepackage{fancyhdr}       % header
\usepackage{graphicx}       % graphics
\usepackage{comment}
\usepackage{longtable}
\usepackage{pifont}% http://ctan.org/pkg/pifont
\usepackage[mode=match]{siunitx}
\usepackage[T1]{fontenc}
\usepackage[utf8]{inputenc}

\usepackage{microtype}

\usepackage{graphicx}
\usepackage[inline]{enumitem}
\setlist{topsep=0pt, leftmargin=*}

\usepackage{algorithm}
\usepackage[noend]{algpseudocode}
\makeatletter
\def\BState{\State\hskip-\ALG@thistlm}
\makeatother

\usepackage{xcolor}
\usepackage{color}
\definecolor{mygreen}{rgb}{0,0.6,0}
\definecolor{mygray}{rgb}{0.85,0.85,0.85}
\definecolor{mymauve}{rgb}{0.58,0,0.82}
\definecolor{mygray}{gray}{0.7}
\usepackage[most]{tcolorbox}
\usepackage{diagbox}

\usepackage{listings, listings-rust}
\usepackage{letltxmacro}
\LetLtxMacro\oldttfamily\ttfamily
\DeclareRobustCommand{\ttfamily}{\oldttfamily\csname ttsize\endcsname}
\newcommand{\setttsize}[1]{\def\ttsize{#1}}%
\setttsize{\small}

\usepackage[inline]{enumitem}
\setlist{topsep=0pt, leftmargin=*}

\usepackage{colortbl}
\newcommand{\lightmidrule}{\arrayrulecolor{black!30}\midrule\arrayrulecolor{black}}

\newcommand{\unitcoder}{UnitCoder\xspace}
\newcommand{\fuzzcoder}{FuzzCoder\xspace}
\newcommand{\unitllama}{UnitLlama\xspace}
\newcommand{\fuzzllama}{FuzzLlama\xspace}
\newcommand{\unitsyn}{UniTSyn\xspace}
\newcommand{\nrepo}{249\xspace}

\newcommand{\fuzzaug}{FuzzAug\xspace}
\newcommand{\dir}[1]{\texttt{#1\slash}}

\usepackage{newpxtext}
\makeatletter
\newcommand{\ie}{\emph{i.e.}\@ifnextchar.{\!\@gobble}{}}
\newcommand{\eg}{\emph{e.g.}\@ifnextchar.{\!\@gobble}{}}
\newcommand{\etc}{etc\@ifnextchar.{}{.\@}}
\makeatother

\NewDocumentCommand{\rscode}{v}{%
  \lstinline[language=rust, basicstyle=\ttfamily]|#1|%
}

\title{\fuzzaug: Data Augmentation by Coverage-guided Fuzzing \\ for Neural Test Generation}

\author{
  Yifeng He,\, Jicheng Wang,\, Yuyang Rong,\, Hao Chen \\
  University of California, Davis \\
  \small{{\{yfhe,\, jicwang,\, chen\}@ucdavis.edu,\, PeterRong96@gmail.com}}
}

\begin{document}
\maketitle

\input{src/abstract}
\input{src/intro}

\input{src/design}

\input{src/experiment}

\input{src/results}

\input{src/related}

\input{src/conclusion}
\input{src/limitations}

\bibliography{main}

\newpage
\appendix
\input{src/appendix}

\end{document}

%% file: src/abstract.tex
\begin{abstract}
	Testing is essential to modern software engineering for building reliable software.
	Given the high costs of manually creating test cases,
	automated test case generation, particularly methods utilizing large language models,
	has become increasingly popular.
	These neural approaches generate semantically meaningful tests that are more maintainable compared with traditional automatic testing methods like fuzzing.
	However, the diversity and volume of unit tests in current datasets are limited, especially for newer but important languages.
	In this paper, we present a novel data augmentation technique, \emph{\fuzzaug},
	that introduces the benefits of fuzzing to large language models 
	by introducing valid testing semantics and providing diverse coverage-guided inputs.
	Doubling the size of training datasets,
	\fuzzaug improves the performances from the baselines significantly.
	This technique demonstrates the potential of introducing prior knowledge from dynamic software analysis
	to improve neural test generation,
	offering significant enhancements in neural test generation.
\end{abstract}

% This enhances the model's ability to embed correct inputs
% that can explore more branches of the function under test.
% with dynamic behaviors of the function under test.
% Our evaluations show that models trained with dataset augmented by \fuzzaug
% increase assertion accuracy by 5\%,
% improve compilation rate by more than 10\%,
% and generate unit test functions with 5\% more branch coverage.

%% file: src/intro.tex
\section{Introduction}
Testing is one of the most important processes in software engineering,
ensuring the quality and reliability of large software applications.
Unit tests are example-based self-assessment tests written and executed by the developer to demonstrate that
the software works correctly as described in the design specification~\cite{runeson2006survey}.
However, despite its importance,
developers do not always contribute new tests due to the difficulty of
identifying which code to test, isolating them as fine-grained units, and finding relevant inputs~\cite{daka2014survey}.
Heuristic-based automatic unit test generation~\cite{pacheco2007randoop,fraser2011evosuite} is one solution to these issues,
% by treading all functions as units under test (focal function),
but the resulting tests are unsatisfactory in
readability, correctness, and diversity of relevant input-output pairs~\cite{panichella2020revisiting}.
% As a result, automatic randomized testing has diverged into
Other popular automatic randomized testing methods,
\eg. fuzzing~\cite{serebryany2016libfuzzer},
often ignores readability and focuses only on generating inputs to find new program behaviors, \ie. new coverage or crashes.
% and property-based testing~\cite{claessen2000QuickCheck},
% which lets developers write formal specifications and only generate random inputs.
However, these randomized testing methods only provide the input that triggers the bug with no valid semantics.
These reported input seeds are usually not as informative as unit test functions in practice~\cite{pbt_in_practice}.
Therefore, finding semantic meaningful test cases correctly and effectively remains an unsolved problem.

More recently, 
people have attempted to overcome these issues by leveraging the power of generative language models~\cite{nie2023TeCo,rao2023catlm, he2024unitsyn}.
Large language models (LLMs) trained on large code corpora can write meaningful programs given text descriptions~\cite{qwen,rozière2023codellama,lozhkov2024starcoder2stackv2}.
Therefore, with sufficiently large code and test datasets,
we expect that LLMs could generate high-quality unit tests to assist human software engineers.

However, testing functions typically occupy a minor fraction of a software repository,
compared with regular functions for software features.
\citet{rao2023catlm} found that in popular Python and Java repositories,
test files comprise fewer than 20\% of all code files.
This deficiency in training data hampers the ability of LLMs to generate practical tests for production environments for two reasons:
\begin{enumerate*}
	\item the imbalance in training data causes the model to miss critical details in the units under test.
	\item the insufficient amount of testing code presents a significant challenge in learning the representations of unit tests adequately.
\end{enumerate*}
Previous work addressed the imbalance issue by aligning code and tests into pairs~\cite{rao2023catlm,he2024unitsyn}.
However, the second issue remains unsolved,
and is further amplified by the trend of switching to newer programming languages for better maintainability and reliability, \eg. redesigning software in Rust.

\begin{comment}
Code-test alignment is an effective approach to solve the first issue,
where CAT-LM~\cite{rao2023catlm} introduced file-level alignment and
UniTSyn~\cite{he2024unitsyn} improved the alignment to a more fine-grained function level,
referred to as function-level code-test pairs. % throughout the rest of this paper.
UniTSyn tackles the second challenge by designing an extendable test collection framework
to gather testing code from repositories in various languages to enhance the diversity and quantity of tests.
Although this approach can build large-scale datasets for test generation,
the size of unit test functions in the dataset is still limited since they have to be created by human developers.
This problem is further amplified by the trend of switching to newer programming languages for better maintainability and reliability.
Newer languages, like Rust, often support important features for security and efficiency,
but have fewer training resources available compared to widely adopted languages like C/C++ and Java.
\end{comment}

A promising strategy to further enhance the existing state-of-the-art unit test datasets is
designing a new specialized data augmentation (DA) method for LLM-based test generation. %~\cite{rebuffi2021data, rubin1987comment}.
In computer vision, data augmentation typically involves applying randomized geometric or color
\emph{transformations} or injecting \emph{random noise} to images in the training set.
However, these methods are unsuitable for programming languages (PLs) due to their formal grammar and strict semantics.
% Limited existing work on data augmentation for code often apply semantic-preserving transformations~\cite{yu2022data},
% including identifier renaming, changing for-loops to while-loops, \etc.
% While these methods enhance the performance of encoder-based code understanding models, 
% they may not directly apply to test generation models,
Limited research~\cite{yu2022data} on DA for PL is not suitable for test generation,
as they do not introduce new test cases that explore the behavior of the program.
Unit test functions provide correct setups to invoke the functions under test (focal functions),
and test inputs are fed to the focal functions to explore their functionality at run-time.
Consequently, a valid data augmentation method for test generation must incorporate semantically meaningful unit test functions,
coupled with randomized yet valid testing inputs tailored to the specific functions under test.

\begin{figure*}[t]
	\centering
	\includegraphics[width=\linewidth]{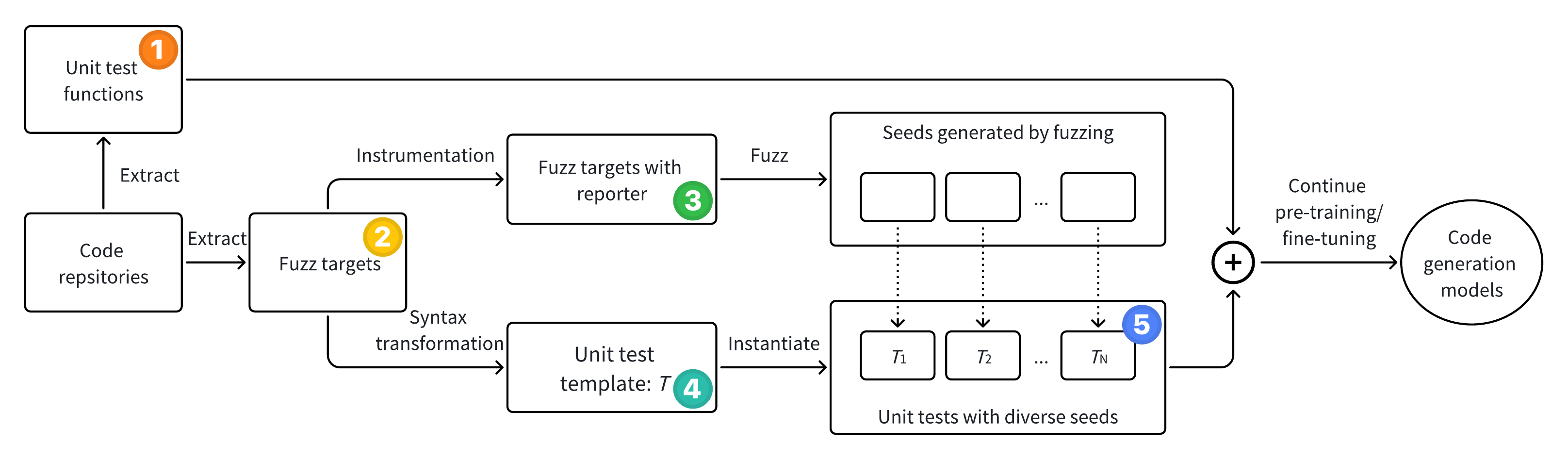}
	\caption{%
		Data Augmentation by fuzzing for neural test generation.
		To construct the augmented dataset, 
		we first extract unit test functions (\autoref{lst:rust:unit_test}) and fuzzing targets (\autoref{lst:rust:fuzz_target}). 
		We instrument each fuzz target with a reporter (\autoref{lst:rust:fuzz_target_reporter}) to collect fuzzing seeds.
		We transform each fuzz target into a unit test template (\autoref{lst:rust:test_template}).
		Finally, we instantiate the templates with valid test inputs to create the augmented training dataset (\autoref{lst:rust:generated_test_func}).
		%  for training code generation models.
        Please refer to \autoref{fig:rust_fuzz_target} for examples of each step.
	}
	\label{fig:fuzzaug}
\end{figure*}

To address these challenges, we propose \emph{\fuzzaug}.
\fuzzaug, as depicted in \autoref{fig:fuzzaug},
is a direct and effective data augmentation technique utilizing fuzzing data to enhance test generation with LLMs.
Fuzzing identifies vulnerabilities in software by randomly generating inputs to trigger new execution paths in software.
These inputs capture the program's runtime behavior and thus can enhance the code understanding capabilities of LLMs~\cite{zhao2023understanding,huang2024code}.
For implementing fuzzing data as a form of data augmentation,
we perform code transformations on fuzz targets in libFuzzer~\cite{serebryany2016libfuzzer} to create new unit test functions.
\fuzzaug nearly doubles the limited amount of testing code
in training datasets and provides a richer diversity of accurate and executable inputs for the focal functions.
Training LLM-based test generation models with \fuzzaug addresses the aforementioned issues
by automatically providing unit test functions with high-quality test inputs.
Thus, \fuzzaug is a novel approach in training practical LLM-based unit test assistance,
enhancing software robustness and maintaining test readability.

To assess the effectiveness of \fuzzaug,
we conducted experiments with three different state-of-the-art 7B open-source code generation models.
% StarCoder2~\cite{lozhkov2024starcoder2stackv2}, CodeLlama~\cite{rozière2023codellama}, and CodeQwen~\cite{qwen}.
Each model was trained on two datasets: on the original \unitsyn~\cite{he2024unitsyn} dataset and its \fuzzaug-augmented counterpart.
All three models trained with \fuzzaug consistently outperformed their counterparts trained on only \unitsyn,
and outperformed the pre-trained/instruction-tuned baseline significantly.
They demonstrated significant improvements in generating accurate test cases (assertions) 
and useful test functions that achieved higher code coverage.
% These findings underscore \fuzzaug's potential to enhance the quality and effectiveness of unit test generation,
% effectively addressing the limitations of existing unit test datasets.

\paragraph{Our contributions.}
\begin{enumerate*}[label=\textbf{\arabic*}.]
	\item We introduce \fuzzaug, a novel data augmentation method 
	      specifically designed for neural test generation LLMs
		  to address the limitations of existing training datasets.
	    % \fuzzaug addresses the critical issue of insufficient testing code in repositories
	    %   and the need for precise and diverse inputs to invoke the focal function.
	    %   To the best of our knowledge, \fuzzaug is the first data augmentation method for neural test generation.

	\item We build and release the Rust version of \unitsyn,
	      % comprising functional-level code-test pairs,
	      aiming at training test generation models for Rust programs.
	      Furthermore, we apply \fuzzaug to this dataset and release the resulting augmented dataset,
	      enhancing its utility for advanced model training.

	\item We validate the efficacy of \fuzzaug by training generative LLMs on the \unitsyn dataset augmented by it.
	      The notable improvement underscores the necessity and advantages of incorporating fuzzing-augmented
	      testing functions into the training corpus, demonstrating the practical benefits of our approach.
\end{enumerate*}

%% file: src/design.tex
\section{Design of \fuzzaug}

\begin{figure*}[t]
	\input{code/fuzz_target}
	\vspace{-1\baselineskip}
	\caption{%
		Simplified examples from \texttt{base64}~\cite{base64} in our collected Rust dataset.
		Each example listing corresponds to one step in \autoref{fig:fuzzaug}.
		Please refer to \autoref{sec:testing_format} for details of unit testing in Rust.
	}
	\label{fig:rust_fuzz_target}
\end{figure*}

\subsection{Challenges}

Generating meaningful test functions as training data for neural test generation models is a complex and critical challenge.
To introduce high-quality random data for training test generation models,
a data augmentation method should satisfy the following requirements:
\begin{enumerate*}
	\item The randomly generated data must be meaningful and valid to the software testing context,
	      \ie, the random data should be able to explore the program's behavior space.
	\item The augmentation modification must provide valid testing semantics in the unit test functions.
	      % keep the syntax and semantic integrity of test functions.
	      As stated by \citet{pacheco2007randoop},
	      unit test functions must correctly parse the random input, set up the state by invoking the focal function,
	      and assert the result of the final call is desired when possible.
\end{enumerate*}

Therefore, designing data augmentation to train test generation models involves creating a sophisticated balance.
On the one hand, introducing sufficient variability to train the models under diverse conditions is essential to generate high-quantity test cases.
On the other hand, maintaining the semantic integrity of augmented test functions is crucial to ensure the validity of training data.
%  the generated tests are 
% valid and applicable in real-world software testing processes. 
This makes the development of \fuzzaug not only challenging but also vital for advancing the capabilities of neural test generation
with language models.

\subsection{Fuzzing for Random Input}

The first requirement ensures that the randomly generated data is beneficial to model training.
High-quality test cases are expected to reflect the behavior of the programs,
which is hard to achieve by data augmentation for natural language data.
To improve the model's ability to generate useful test cases,
the data augmentation method needs to be aware of the program's structure and behavior.

\paragraph{Fuzzing.}
Fuzzing is a widely used software testing method that generates inputs randomly to explore unseen program behaviors~\cite{zeller2019fuzzing}.
Coverage-guided fuzzing can be summarized as a four-stage loop consisting of input generation, program execution, behavior monitoring, and input ranking. % as shown in \autoref{fig:fuzzing}.
First, the program is executed with a given input.
During execution, the program's dynamic behavior, particularly branch coverage, is monitored to collect coverage information.
If a new behavior is observed, the triggering input is saved in a seed queue and prioritized for next round of mutation; otherwise, it is discarded.
Finally, the mutator modifies the input for the next cycle to explore new behaviors.
Various mutation, behavior monitoring, seed scheduling strategies
have been studied to enhance the quality of input seeds during fuzzing~\cite{aflfast,aflgo,chen2018angora,she2019neuzz},
and are integrated to the modern fuzzers like LibFuzzer~\cite{serebryany2016libfuzzer}.

Fuzzers select input seeds by executing the programs,
these inputs embed the program's dynamic behavior and are thus able to discover bugs and vulnerabilities in the program.
Previous studies~\cite{zhao2023understanding,huang2024code} show that fuzzing input-output pairs are helpful
for language models to understand programs.
Therefore, we argue that random inputs generated by fuzzers are also suitable to contribute to randomized mutation
for testing function data augmentation.
Thus, this first requirement is satisfied by engaging fuzzing in the data augmentation process.

% Fuzzing can be conducted on different levels of the software.
% % AFL++~\cite{fioraldi2020afl_pp} feeds inputs to whole programs and tests the program entirely.
LibFuzzer~\cite{serebryany2016libfuzzer} %, on the other hand,
allows users to define custom fuzz targets to specify the most important functions as entry points for testing.
We select libFuzzer for its function-level fuzzing feature to ensure syntax correctness when invoking the corresponding focal function.
If we can compile and run the fuzz target successfully,
we are confident that the testing code is valid training data for the language model.
Therefore, the validity of \fuzzaug is guaranteed.
To collect inputs with the program's dynamic behavior from the fuzzing loop,
we instrument a \texttt{reporter} to each fuzz target as shown in \autoref{fig:fuzzaug}.
% We use \texttt{reporter} to denote the reporting process in the figure for simplicity.
After all the fuzz targets in the project are instrumented,
we start the fuzzing loop for each target and save the reported inputs
as a randomly generated portion of our data augmentation process.

\subsection{Unifying Code Representation} \label{sec:transformation}

For code generation with causal language modeling,
valid and complete training data
with appropriate semantics within the tokens is beneficial.
Therefore, to avoid any distribution shift between unit test functions and data augmentation,
we cannot append inputs generated by fuzzing to training data directly
due to the distinct representations between raw fuzzing inputs and meaningful unit test functions.
Fuzzers treat all inputs as bytes and apply byte-level random mutations, for example, bit-flip.
Previous work on using fuzzing data for code understanding tasks decodes the raw inputs
into strings and append the inputs to the program~\cite{zhao2023understanding}
or uses different language modeling loss functions for two kinds of data~\cite{huang2024code}.
However, these approaches do not apply to generative models,
so we need to design a different representation for fuzzing data.
% to train with CLM.

\begin{comment}
Since we aim to teach language models to write better unit tests, % with accurate test cases and complete test function with high coverage,
the augmented data must be in valid syntax and carry test semantics.
The validity of syntax is guaranteed by executing the fuzz targets.
To compile the fuzzers successfully,
there must exist no syntax error in the focal functions and fuzz targets.
In addition, the implementation of the fuzz targets naturally embeds the testing semantics.
Fuzz targets, as shown in \autoref{lst:rust:fuzz_target},
are perfect examples of testing the focal functions under given input data.
Therefore, we can transform the fuzz targets into unit test templates that keep both
syntax validity and semantic integrity of test functions.
\end{comment}

We implement a syntax transformation in the compiler frontend to obtain valid new test functions to keep testing semantics.
We compiled these candidates (\autoref{lst:rust:fuzz_target}) into Abstract Syntax Trees (ASTs) and extracted the function bodies from each AST
using \texttt{proc\_macro}~\cite{proc-macro2} and \texttt{syn}~\cite{syn}.
Then we rewrite the macro for fuzz targets into valid function definitions
with the \rscode{#[test]} attribute on top to help test discovery (\autoref{lst:rust:test_template}).
We call the result of syntax transformation \emph{test template}.
We demonstrate a fuzz target and its transformed test template in \autoref{fig:fuzzaug}.
These test templates are stored for actual data augmentation at a later stage.

\subsection{Fuzz Augmentation}

\begin{algorithm}[t]
	\caption{Fuzzing as Data Augmentation}
	\label{alg:fuzzaug}
	\small
	\input{code/fuzzaug.tex}
\end{algorithm}

% Since fuzzing is high-throughput, we record an exceeding amount of inputs during the process.
To ensure the quality of the augmented data,
we employed an input selection algorithm as shown in \autoref{alg:fuzzaug}.
Raw inputs collected from fuzzing have two drawbacks.
First, there will be repeated or overlapping inputs collected from fuzzing.
Fuzzing applies mutation on inputs that explore new paths in the program.
Therefore, consecutive inputs differ only in small parts,
which should be avoided.
% These characteristics of fuzzing data will cause the trained model to generate repeated test cases.

Second, since the input data are generated randomly by libFuzzer~\cite{serebryany2016libfuzzer},
the token length for those inputs can be excessively long.
This behavior happens especially commonly when the input type is a vector or long number (\rscode{i64}, \rscode{f64}, etc)
since the length of the vectors or numbers is not a problem for fuzzing.
% where inputs are stored in random access memory or disks.
However, for generative models, the acceptable token length is much smaller,
so such long inputs will harm the performance of the model.
To overcome the aforementioned issues,
we designed our selection algorithm to first shuffle the inputs and then sample the desired inputs within a given length.
Our algorithm samples $N$ fuzzing inputs that satisfy the requirements
to instantiate the test templates for unique data augmentation (\autoref{lst:rust:generated_test_func}).

\begin{comment}
Then for each transformed test template,
we duplicate it for $N$ copies.
Finally, we instantiate each copy of the test template with valid selected fuzzing inputs
to generate valid new unit test functions with diverse inputs as data augmentation.
Thus, for every selected fuzz input, \fuzzaug generates a unique unit test function.
\end{comment}

%% file: code/fuzz_target.tex
\begin{minipage}{\linewidth}
	\begin{subfigure}[t]{0.49\linewidth}
		\begin{lstlisting} [
            language=Rust, 
            caption={Unit test function extracted from repository}, 
            label=lst:rust:unit_test,
        ]
#[test]
fn encode_all_bytes_url() {
    let bytes: Vec<u8> = (0..=255).collect();
    assert_eq!(
        "...", // expected result 
        &engine::GeneralPurpose::new(&URL_SAFE, PAD).encode(bytes)
    );
}
        \end{lstlisting}
	\end{subfigure}
	\begin{subfigure}[t]{0.49\linewidth}
		\begin{lstlisting} [
            language=Rust, 
            caption={Fuzz target extracted from repository}, 
            label=lst:rust:fuzz_target,
        ]
#![no_main]
#[macro_use] extern crate libfuzzer_sys;
extern crate base64;
use base64::*;
mod utils;
fuzz_target!(|data: &[u8]| {
    let engine = utils::random_engine(data);
    let _ = engine.decode(data);
});
        \end{lstlisting}
	\end{subfigure}
\end{minipage}
\begin{minipage}{\linewidth}
	\begin{subfigure}[t]{0.49\linewidth}
		\begin{lstlisting} [
            language=Rust, 
            caption={Fuzz target instrumented with reporter}, 
            label=lst:rust:fuzz_target_reporter,
        ]
fuzz_target!(|data: &[u8]| {
    report(data); // example reporter
    let engine = utils::random_engine(data);
    let _ = engine.decode(data);
});
        \end{lstlisting}
	\end{subfigure}
	\begin{subfigure}[t]{0.49\linewidth}
		\begin{lstlisting} [
            language=Rust, 
            caption={Test template transformed from fuzz target}, 
            label=lst:rust:test_template,
        ]
#[test]
fn test_template() {
    let data = []; // example template
    let engine = utils::random_engine(data);
    let _ = engine.decode(data); }
        \end{lstlisting}
	\end{subfigure}
\end{minipage}
\begin{subfigure}[t]{\linewidth}
	\begin{lstlisting}[
        language=rust, 
        caption={Unit test function instantiated from test template with a seed generated by fuzzing}, 
        label=lst:rust:generated_test_func,
      ]
#[test]
fn test_1() {
    let data = [3,44,12,3,21,2,255,12,4,34,12,4,12,3];  // example recorded test input
    let engine = utils::random_engine(data);
    let _ = engine.decode(data); }
      \end{lstlisting}
\end{subfigure}

%% file: code/fuzzaug.tex
\begin{algorithmic}[1]
\Function{\fuzzaug}{$repo$, $N$, $L$, $timeout$} \label{lst:func:fuzzaug}
    \\\Comment{$repo =$ repository to apply \fuzzaug}
    \\\Comment{$N =$ number of training examples to generate}
    \\\Comment{$L =$ maximum input length for collection}
    \\\Comment{$timeout =$ maximum allowed fuzzing time}
    \State $dataset_{\text{aug}} \gets$ []
    \ForAll{$t \in$ \textproc{GetFuzzTarget}($repo$)}
        \State $t' \gets$ \textproc{ReporterInstrumentation}($t$)
        % \Comment{Collect raw fuzzing inputs}
        \State $inputs \gets$ \textproc{Fuzz}($t'$, $timeout$)
        \State $inputs' \gets$ \textproc{Filter}($\lambda x:$ \textproc{len}($x$) $< L$, $inputs$)
        \State $selected \gets$ \Call{Sample}{$N$, $inputs'$}
        \State $templates \gets$  \Call{SyntaxTransformation}{$t$}
        \State $aug \gets$ \Call{Instantiate}{$templates[:N]$, $selected$}
        % \State $templates \gets$ \textproc{Take}($N$, \Call{SyntaxTransformation}{$t$})
        \State $dataset_{\text{aug}} \gets dataset_{\text{aug}} + aug$
    \EndFor
    \State \Return $dataset_{\text{aug}}$
\EndFunction
\end{algorithmic}

%% file: src/experiment.tex
\section{Experimental Setup} \label{sec:experiment}

\subsection{Data Collection}
\begin{table}[t]
	\centering
	\begin{adjustbox}{max width=\columnwidth}
		\input{tabels/data_stat.tex}
	\end{adjustbox}
	\caption{%
		Dataset statistics.
		Unit tests: the base dataset we collected from code repositories using \unitsyn~\cite{he2024unitsyn}.
		% This dataset is used to train UnitCoder, UnitQwen, and UnitLlama in \autoref{sec:experiment}.
		Fuzz: the dataset we transformed from fuzz targets using \autoref{alg:fuzzaug}, where $N = 40$.
		Augmented dataset: the combination of unit tests and fuzz.
		% which is used to train the fuzz models in \autoref{sec:experiment}.
	}
	\label{tab:data_stat}
\end{table}

We chose Rust language to conduct this research for three reasons.
First, Rust projects are highly structured with \dir{src},
\dir{tests}, and \dir{fuzz} directories on the top level.
With the \texttt{cargo} package manager,
we can build and run the project without solving dependency issues.
Second, the Rust compiler has built-in support for
unit testing and fuzzing,
so collecting unit tests and fuzzing data is straightforward.
Third, Rust's syntax for libFuzzer passes a closure to a predefined macro,
so we can apply syntax transformation described in \autoref{sec:transformation} to the fuzz targets.
Rust is one of the most popular languages for security-critical software,
and yet is new compared to older languages like C/C++,
further lighting the necessary for effective data augmentation.
We follow \unitsyn~\cite{he2024unitsyn} to collect the training data from open-source repositories on GitHub.
% transform fuzzers to unit test format.
% This eases the problem that the model needs to learn
% two different representations for the fuzzing seeds and unit test functions.%  two kinds of testing data.

\begin{comment}

\paragraph{Repository and package mining.}
% We follow previous work~\cite{husain2019codesearchnet,rao2023catlm,he2024unitsyn}
% to collect open-source repositories from GitHub.
We follow \unitsyn~\cite{he2024unitsyn} to select Rust repositories
based on stars and the date of the last commit.
% We require the repositories to be popular (with more than ten stars)
% and under active development (with new commits after January 1st, 2020).
% We also de-duplicate the dataset by filtering out repositories that are archived, forked, or mirrored.
In addition to the requirements in \unitsyn,
we also require the repository to contain pre-defined unit tests and fuzzers in order to apply \fuzzaug.
For this purpose,
we check the presence of \dir{tests} and \dir{fuzz} directories in the root of the project.
In total, we find \nrepo suitable repositories to build our dataset.
\end{comment}

\paragraph{Unit test collection.} \label{section:data_collection}

Different from previous work training on file-level code-test pairs \citep{rao2023catlm},
we follow previous work~\cite{nie2023TeCo,he2024unitsyn} to collect our training data as function-level code-test pairs
since it suits our data augmentation method.
We implement the Rust hook for the \unitsyn~\cite{he2024unitsyn}
based on the \rscode{#[test]} attribute on top of the Rust unit test functions.
To find the call to the focal function,
since assertion in Rust is a macro instead of a keyword or function as in \unitsyn,
we extend the framework to handle this marco special case.
From the downloaded repositories, we found \num{14633} calls to the focal functions in the unit tests,
and collected \num{7881} focal-test pairs as training data.

\paragraph{Augmented test collection.}
We chose LLVM libFuzzer \citep{serebryany2016libfuzzer} to utilize the pre-defined fuzz targets in the code repositories. % for it's function-level fuzzing feature.
For Rust, libFuzzer is supported as \texttt{cargo-fuzz}.
%  mentioned in the previous sections.
We instrumented each fuzz target in the repository to report the input fuzzing data.
We transform the body of the fuzz target macro to an equivalent unit test template,
as described in \autoref{fig:fuzzaug}.
We fuzzed all targets for one minute following previous work~\cite{zhao2023understanding,huang2024code} on fuzzing for code understanding.
All fuzzing processes are performed on a server with dual 20-core, 40-thread x86\_64 CPUs and 692 GB of RAM.
%then substituted the reported data back to the generated unit test function template for training.
Out of the \nrepo repositories we downloaded, 179 of them can be compiled successfully for fuzzing.
For the main experiments,
we set $N = 40$ so that the augmented data is at the same scale as the original unit test dataset,
and explore the effects of scaling $N$ later in \autoref{sec:scaling}.
We collected in total of \num{6811} additional code-test pairs generated by \fuzzaug.
The statistics of the collected unit test dataset and data augmentation are summarized in \autoref{tab:data_stat}.

\subsection{Baseline Models} \label{sec:training}
\begin{table}[t]
	\centering

	\begin{adjustbox}{max width=\columnwidth}
	\input{tabels/models.tex}

	\end{adjustbox}
	\caption{%
		Our model selection for evaluation. \\
		Base Model: names of the baseline models used for applying the fine-tuning methods.
		% Type: pre-trained~(PT), instruction-tuned~(PT).
		% \unitsyn: corresponding models fine-tuned using the \unitsyn~\cite{he2024unitsyn} dataset.
		% \fuzzaug: corresponding models fine-tuned using \unitsyn augmented by \fuzzaug.
	}
	\label{tab:models}

\end{table}

We select three baselines to evaluate \fuzzaug.
% StarCoder2~\cite{lozhkov2024starcoder2stackv2}, CodeQwen1.5~\cite{qwen}, and CodeLlama~\cite{rozière2023codellama}.
StarCoder2~\cite{lozhkov2024starcoder2stackv2} is the successor of UniTSyn's base model SantaCoder~\cite{allal2023santacoder}.
% is a state-of-the-art open-source code generation model with 7B parameters.
% we use it for a fair comparison against \unitsyn with no data augmentation.
We follow EvalPlus~\cite{evalplus} to select the best-performing 7B code generation model CodeQwen1.5~\cite{qwen}.
Finally, we experiment on CodeLlama~\cite{rozière2023codellama} to compare against its instruction-tuned baseline.
The complete model selection and naming are in \autoref{tab:models}.
% which is powerful enough to generate valid code and efficient enough to evaluate the augmented data.
Our training details are in \autoref{sec:training_details}.

\subsection{Research Questions}

To evaluate \fuzzaug,
we structure our experiments around the following research questions
on the quality of generated unit tests:

\paragraph{RQ.1. Can \fuzzaug improve the accuracy of generated test cases?}
Software testing aims to discover hidden bugs in the code.
The prerequisite of this aim is to have \emph{accurate} test cases,
where the generated input and output to the focal function match with the ground truth.
% Therefore, measuring the accuracy of generated test cases is essentially
Therefore, accuracy of generated test cases is an essential metric for software testing.
% to evaluate if the language model is useful in software testing.
Generating accurate test cases %the correct input-output of the focal function
requires the model to learn both the semantics and runtime behavior of the focal function,
which is challenging for language models~\cite{gu2024cruxeval}.
% Predicting the correct inputs and corresponding outputs is challenging for language models~\cite{gu2024cruxeval},
% so it is an important direction for models to improve on.
We follow previous work~\cite{chen2023codet,he2024unitsyn} to extract the first 10 generated test cases
to examine their standalone correctness.
We compile and execute these test cases against the ground truth focal function independently.
% Let $N$ denote the number of generated test cases, $n$ denote the number of test cases that can be compiled,
% and $m$ denote the number of test cases that can be executed with no error,
% we evaluate $\text{Assert. CR} = \nicefrac{n}{N}$ and 
% $\text{Acc} = \nicefrac{m}{N}$.

\paragraph{RQ.2. Can \fuzzaug improve the validity and completeness of generated unit tests?}
Accurate assertions are essential for unit testing,
while completeness and validity are necessary for generated test functions to be practical.
A generated test function is \emph{valid} if it can be compiled and executed. % without over-complex post-processing.
On the other hand, a test function is \emph{complete} if it can cover all of the branches of the focal function.
% while comprehensive coverage on the focal functions \emph{enhances their quality}.
Therefore, we follow \unitsyn to use the compile rate of the whole generated unit test functions and branch coverage on the focal functions to
check the validity and completeness of the generated unit test functions.
% To evaluate the coverage fairly,
% we only apply some basic post-process (\autoref{sec:eval-setup}) on the generated functions,
% run the entire test function,
% and record the branch coverage using \texttt{grcov}~\cite{grcov}.
We use \texttt{grcov}~\cite{grcov} to measure the branch coverage.
% Using branch coverage as a metric for neural test coverage allows
% us to better evaluate the usefulness of the test generation models and of \fuzzaug.
% our data augmentation to improve these models.

\paragraph{RQ.3. Can \fuzzaug generalize to other models?}
Data augmentation is a training-time technique that should improve the performance of all models in the same task.
% The same augmentation method should produce similar performance improvement trends
% on different models.
% To evaluate the generalizability of \fuzzaug,
% we apply the same training and evaluation setting for training three different models
% and report the effects of \fuzzaug.

\paragraph{RQ.4. The effect of further scaling \fuzzaug.}
% In previous experiments, we set $N=40$ to ensure that 
% the augmented data remained on the same scale as the original unit test dataset.
% However, 
It is possible to further scale-up \fuzzaug,
so we explore the effects of hyperparameter $N$.

\subsection{Evaluation Setup} \label{sec:eval-setup}

% \subsubsection{Evaluation datasets}
\paragraph{Benchmark dataset.}

We follow \unitsyn to evaluate the models on HumanEval-X \cite{zheng2023codegeex},
a hand-crafted  benchmark for code generation tasks that contains Rust.
% containing problems in popular languages including Rust.
HumanEval-X has 164 different problems,
where each of them is composed of
description prompt in natural language, function declaration (header), canonical solution (ground truth implementation), and unit test function.
% HumanEval-X was designed for code generation and code translation in different languages,
% but is also useful for test generation.
We follow UniTSyn to use the canonical solution as the focal function,
and let the model generate the corresponding test function.

\paragraph{Prompts.}
% For checking the accuracy of generated assertions,
We follow \citet{chen2023codet}
to guide the language models in generating assertions (\autoref{lst:prompt}).
We use natural language ``Check the correctness of \textasciigrave\texttt{function\_name}\textasciigrave'' in comments
to instruct the model to complete the test function.
We guide the generation of assertions by providing
the language-specific assert keyword and the incomplete invocation of the focal function.
% as shown on Line~\autoref{lst:assertion_guide}.
We allow the model to predict at most 1024 new tokens for the synthesized assertions for all models.
We set the generation temperature to 1 for all the models to encourage output diversity.
We concatenate the import statements, the focal function implementation,
the natural language instruction in the comment, and the test header
together as the import prompt to the language model.

\input{code/prompt}

\paragraph{Post-processing.}
We avoid overly intricate processing of the generated test functions to keep our evaluation results faithful.
% We only apply the most basic post-processing to make the generated test function complete.
We first count the number of the curly brackets.
% If the number of opening brackets matches the number of closing brackets,
% we consider the generated test function as complete.
If the numbers do not match,
we check if the last generated line ended with a semicolon
to see if the last line is complete.
If not, we remove that line.
Then we add the missing closing curly brackets to complete the generated test.

%% file: tabels/data_stat.tex
\begin{tabular}{lrrrr}
	\toprule
	Dataset    & \# Repo & \# Focal    & \# Pairs    & \# Tokens \\
	\midrule
	Unit tests & 249     & \num{14633} & \num{7881}  & 2.5M      \\
	Fuzz       & 179     & \num{14790} & \num{6811}  & 2.2M      \\
	\midrule
	All        & 249     & \num{29423} & \num{14692} & 4.7M      \\
	\bottomrule
\end{tabular}

%% file: tabels/models.tex
\begin{comment}
\begin{tabular}{l|ccc}
	\toprule
	\diagbox{Method}{Base Model} & StarCoder2 (PT) & CodeQwen1.5 (PT) & CodeLlama (IT) \\
	\midrule
	\unitsyn                     & UnitCoder       & UnitQwen         & UnitLlama      \\
	% \midrule
	\fuzzaug                     & FuzzCoder       & FuzzQwen         & FuzzLlama      \\
	\bottomrule
\end{tabular}
\end{comment}
\begin{tabular}{ccccc}
	\toprule
	         & \multicolumn{3}{c}{Base Model} &                                     \\
	\cmidrule(l){2-5}
	Method   & StarCoder2                 & CodeQwen1.5  & CodeLlama & \\
	\midrule
	\unitsyn & UnitCoder                      & UnitQwen         & UnitLlama        \\
	\fuzzaug & FuzzCoder                      & FuzzQwen         & FuzzLlama        \\
	\bottomrule
\end{tabular}

%% file: code/prompt.tex
\begin{lstlisting} [
    belowskip=-2\baselineskip,
    float,
    language=Rust, 
    caption={Example prompt used for test generation. Import statements are removed for simplicity.}, 
    label=lst:prompt,
]
fn has_close_elements(numbers: Vec<f32>, threshold: f32) -> bool { ... }
// Check the correctness of  `has_close_elements`
#[cfg(test)]
mod tests {
    use super::*;
    #[test]
    fn test_has_close_elements() {
        assert_eq!(has_close_elements( (*@ \label{lst:assertion_guide} @*)
\end{lstlisting}

%% file: src/results.tex
\section{Evaluation Results}

We report our experimental results on the performance of neural test generation in this section.
We categorize the models into three groups:
pre-trained (PT), instruction-tuned (IT), and fine-tuned (FT) models.
PT and IT models are the baselines,
while FT models are further trained with \unitsyn and \fuzzaug.

\subsection{Test Case Correctness}

We follow CodeT~\cite{chen2023codet}
% use a Rust-specific assert macro and an incomplete call to the focal function
to guide the language models in generating independent test cases (assertions).
Since the assertions are independent,
we can parse them and evaluate each one of them individually.
We present the evaluation results in \autoref{tab:assert}.
% In \autoref{tab:assert},
% We present the accuracy of the models under columns ``Assert. CR'' for the compile rate of individual assertions,
% and ``Acc'' for the accuracy score of the individual assertions.
% Models trained with \fuzzaug show a significant improvement in both metrics over the baselines,
% and outperform models trained on \unitsyn.
%
% For assertion compile rate on StarCoder2,
% applying \fuzzaug leads to a \emph{+6.89\%} increase over the baseline StarCoder2,
% and a \emph{+5.25\%} increase over \unitcoder.
% We also observe a \emph{+11.10\%} and \emph{+4.88\%} increase over CodeLlama and UnitLlama, respectively.
Notably,
CodeQwen1.5 is the strongest model in this assertion compile rate evaluation,
where we observe an increase of \emph{+14.38\%} over CodeQwen1.5
and \emph{+7.37\%} over UnitQwen.
For assertion accuracy,
% \fuzzaug also shows a significant improvement over the baselines.
% On StarCoder2, 
% \fuzzaug achieves a \emph{+3.67\%} increase over the baseline,
% and a \emph{+2.51\%} increase over \unitcoder.
% For CodeLlama, 
% we observe a \emph{+4.94\%} increase over the baseline
% and a \emph{+2.37\%} increase over UnitLlama.
% CodeQwen1.5 is still the strongest model in accuracy.
We observe a \emph{+10.49\%} increase over CodeQwen1.5
and a \emph{+6.16\%} increase over UnitQwen.

% \begin{tcolorbox}[
% 		colback=black!5!white,
% 		colframe=black!75!white,
% 		title=RQ.1. Can FuzzAug improve the accuracy of generated test cases?,
% 		left=0pt,
% 		right=0pt,
% 		top=0pt,
% 		bottom=0pt
% 	]
% \end{tcolorbox}

\input{tabels/assert.tex}

\subsection{Test Validity and Completeness} \label{sec:result:cov}

To evaluate if \fuzzaug can help the model generate valid unit test functions,
we evaluate the generated unit test functions without extracting the individual assertions.
% we removed the aforementioned assertion-guide in the prompt and evaluate the result on the function level.
Results for this experiment are shown in \autoref{tab:func}. 
% under columns ``Func. CR'' for function compile rate of the generated unit test functions.
For whole test function compile rate, 
\fuzzaug also shows stable improvements on all models.
% \fuzzaug improves StarCoder2 by \emph{+13.83\%} and
% outperforms UnitCoder by \emph{+11.39\%}.
% On CodeLlama,
% we observe an increase of \emph{+17.07\%} increase over UnitLlama,
% and \emph{+7.93} over \unitsyn.
% Since CodeQwen1.5 is already a stronger model,
% we observe less improvement,
% which are \emph{+4.88\%} over CodeQwen1.5 and \emph{+12.80\%} over UnitQwen.
On the strongest model, CodeQwen1.5,
we observe an increase of \emph{+4.88\%} over CodeQwen1.5 and \emph{+12.80\%} over UnitQwen.

\fuzzaug also improves the average branch coverage consistently.
% \fuzzaug improves StarCoder2 by \emph{+7.21\%} and
% outperforms UnitCoder by \emph{+5.17\%}.
% % On CodeLlama,
% FuzzLlama outperforms CodeLlama by \emph{+3.77\%} and UnitLlama by \emph{+3.28\%}.
For CodeQwen1.5,
we observe an increase of \emph{+3.73\%} over CodeQwen1.5 and \emph{+3.87\%} over UnitQwen.
Achieving high branch coverage is a hard task for LLMs,
as it requires deep understanding and reasoning ability
over the function's control flow.
For reference,
even with known overfitting issues~\cite{jain2024livecodebench},
GPT-4 can only achieve an average branch coverage of 47.94\%.

\input{tabels/func.tex}

\subsection{Generalizability of \fuzzaug}

Useful data augmentation methods should work on different models.
We fine-tune three different models with \fuzzaug and evaluate their performance,
where all models trained with \fuzzaug show improvements over the baseline pre-trained models and \unitsyn.

\subsection{Scaling \fuzzaug} \label{sec:scaling}

We explore the effects of scaling \fuzzaug to construct larger training datasets. 
To assess the impact of varying amounts of fuzzing inputs, 
we train models with  $N=40,60,80,100$ fuzzing samples for this experiment.

As shown in Appendix~\autoref{fig:scaling}, the impact of scaling \fuzzaug is not consistent across models. In particular, for the stronger base model CodeQwen1.5, 
increasing $N$ does not lead to significant changes. 
Conversely, for weaker base models, scaling $N$ improves both assertion accuracy and compile rate.
When evaluating the test function compile rate,
both FuzzLlama and FuzzCoder exhibit a positive correlation with increasing $N$. 
Additionally, FuzzLlama's accuracy improves with larger $N$, 
while other metrics show no clear trend.

The results suggest that dataset size alone is not the primary factor influencing model performance. 
Instead, the quality of data augmentation,
driven by the test semantics of the fuzz targets and coverage-guided inputs,
plays a more crucial role. 
Therefore, we recommend selecting $N$ at a scale comparable to the original training dataset, which should be enough.

%% file: tabels/assert.tex
\begin{table}[t]
	\centering
	\begin{tabular}{lcccc}
		\toprule
		Model       & Type    & Assert. CR     & Acc            \\
		\midrule
		StarCoder2  & PT      & 64.09          & 31.83          \\
		\unitcoder  & FT      & 65.73          & 32.99          \\
		\fuzzcoder  & FT      & \textbf{70.98} & \textbf{35.50} \\
		\lightmidrule
		CodeLlama   & IT      & 64.57          & 32.13          \\
		% CodeGemma  & IT   & 50.24          & 29.94          \\
		\unitllama  & FT      & 70.79          & 34.70          \\
		\fuzzllama  & FT      & \textbf{75.67} & \textbf{37.07} \\
		\lightmidrule  % 100 steps
		CodeQwen1.5 & PT      & 66.52          & 41.71          \\
		UnitQwen    & FT      & 73.54          & 46.04          \\
		FuzzQwen    & FT      & \textbf{80.91} & \textbf{52.20} \\
		% \midrule
		% \g{GPT-4}   & \g{API} & \g{95.53}      & \g{75.04}      \\
		\bottomrule
	\end{tabular}
	\caption{
		Accuracy of tests generated by LLMs. The best results are highlighted in bold.
		Assert. CR: the compile rate of the individual assertions.
		Acc: accuracy of individual assertions.
	}
	\label{tab:assert}
\end{table}

%% file: tabels/func.tex
\begin{table}[t]
	\centering
	\begin{tabular}{lccc}
		\toprule
		Model      & Type    & Func. CR       & Cov            \\
		\midrule
		StarCoder2 & PT      & 45.73          & 9.88           \\
		\unitcoder & FT      & 48.17          & 11.92          \\
		\fuzzcoder & FT      & \textbf{59.56} & \textbf{17.09} \\
		\lightmidrule
		CodeLlama  & IT      & 54.88          & 15.75          \\
		% CodeGemma  & IT   & 71.95          & 17.02           \\
		\unitllama & FT      & 64.02          & 16.23          \\
		\fuzzllama & FT      & \textbf{71.95} & \textbf{19.52} \\
		\lightmidrule  % 100 steps
		CodeQwen   & PT      & 68.29          & 20.90          \\
		UnitQwen   & FT      & 60.37          & 20.76          \\
		FuzzQwen   & FT      & \textbf{73.17} & \textbf{24.63} \\
		% \midrule
		% \g{GPT-4}  & \g{API} & \g{93.90}      & \g{47.94}      \\
		\bottomrule
	\end{tabular}
	\caption{Evaluations of usefulness of generated unit tests.
		Func. CR: the compile rate of generated unit test functions.
		Cov: the average branch coverage of generated unit test functions on the focal functions.
	}
	\label{tab:func}
\end{table}

%% file: src/related.tex
\section{Related Work}

\subsection{Fuzzing}

Fuzz testing~\cite{zeller2019fuzzing}, or fuzzing,
is a popular execution-based dynamic testing technique with randomized inputs in various software domains~\cite{rong2020int,chen2023hopper,rong2025irfuzzer}.
Fuzzing aims to generate a set of inputs based on the provided set of seeds to achieve high code coverage.
% The fuzzing process can be described as a coverage-guided input modification method.
The fuzzer uses behavior monitoring to find inputs with high branch coverage
and favors those inputs for future input generation~\cite{chen2018angora,she2019neuzz,rong2024valkyrie}.
% Behavior monitoring enables the fuzzer to focus on exploring new program branches 
% that are not yet tested to detect unseen security issues like memory errors.
% AFL++~\cite{fioraldi2020afl_pp} and libFuzzer \cite{serebryany2016libfuzzer}
% are two of the most popular fuzzers that are widely adopted in testing real-world software.
LibFuzzer~\cite{serebryany2016libfuzzer} is integrated into the LLVM compiler infrastructure~\cite{lattner2004llvm},
% to test projects in LLVM-compiled languages,
and can also be used in other mainstream languages~\cite{jazzer,google/atheris}.
% When using libFuzzer, the tester provides a fuzz target to parse the random input and execute the target function,
% providing the testing semantics for the fuzzer.

\begin{comment}
% Coverage-guided grey-box fuzzing validates the software by repeatedly generating inputs for
% the software to execute and prioritize the inputs that reach new paths in the program.
% One of the key components is to determine if the input seed triggers new and interesting behaviors.
% Several related work have proposed methods to improve fuzzing in this direction,
% including using deterministic~\cite{rong2024valkyrie} and gradient-based~\cite{chen2018angora, she2019neuzz} techniques.
\end{comment}

\paragraph{Fuzzing for machine learning.}
Inputs generated by coverage-guided fuzzing can benefit language models in understanding programs,
as they contain information about the program's dynamic behavior~\cite{zhao2023understanding,huang2024code}.
Fuzzing was also adopted as a data augmentation tool to improve the robustness of neural networks~\cite{gao2020fuzz}.
% SenSei~\cite{gao2020fuzz} introduced fuzzing to improve the robust generalization of DNNs in training.
% Their key insight is that guided search like fuzzing and genetic programming can be more effective in finding optimal variants
% compared to other DA-produced random variants as described by \citet{krizhevsky2012imagenet},
% but is not related with improving LLMs to generate better unit tests.
% The authors also pointed out that SenSei can help the training process
% by skipping training data that is considered ``not interesting'' by fuzzing.
% This work mainly focuses on improving the robustness of neural networks,
% and is not a work to improve generative language models and neural test generation.
% Although both incorporated fuzzing to mutate the training dataset,
% SenSei and our work \fuzzaug are fundamentally different in approach and application.

\subsection{Test Generation via LLMs}

Using especially LLMs to generate test cases
is a new trend in automatic software testing.
This method is referred to as \emph{neural test generation}.
The direct approach toward neural test generation is to instruct
pre-trained code generation LLMs~\cite{rozière2023codellama,lozhkov2024starcoder2stackv2},
or foundation models~\cite{achiam2023gpt,schäfer2023empiricalevaluationusinglarge,tang2024chatgpt}.
The other approach is to train test-specific models that are specialized in generating test cases or test functions~\cite{watson2020learning_assert,tufano2021AthenaTest,dinella2022TOGA,alagarsamy2023a3test}.
The more recent work~\cite{nie2023TeCo,rao2023catlm,he2024unitsyn} proposed to train the test generation model
on \emph{aligned} data that includes the correspondence between the unit test and the function under test (focal).

%% file: src/conclusion.tex
\section{Conclusion}
We developed \fuzzaug, a data augmentation method for unit test function generation.
\fuzzaug combines the advantages of coverage-guided fuzzing
and generative large language models to generate tests that are not only
semantically meaningful but also strategically comprehensive.
We applied \fuzzaug to fine-tune three state-of-the-art 7B open-source code generation models
to demonstrate the effectiveness of \fuzzaug.
% Although our approach is efficient and can be automated,
% it requires the crate's authors and contributors to define the fuzzers manually.
% Therefore, 
We collect our experimental dataset on Rust crates that have pre-defined fuzzers
as a Rust extension to \unitsyn.
Our method can be generalized to all languages that OSS-Fuzz supports with slight modifications. 
Our results show the effectiveness of employing dynamic program analysis to generate high-quality inputs
to augment the code corpus in training language models.
We believe \fuzzaug can spur the development of unit test generation by large language models
and contribute to the field of AI for software engineering and testing.
Our code and artifacts are available anonymous (\href{https://doi.org/10.5281/zenodo.14873588}{link}),
and will be publicly available after publication.

%% file: src/limitations.tex
\section*{Limitations and Future Work}

In this section,
we discuss the potential concerns of our design and limitations.
We structure each concern we foresaw and the discussion of them as subsections.

\subsection*{Applying to Different Languages} \label{sec:diff_lang}

On the high level,
fuzzing is a programming language agnostic testing approach.
LibFuzzer is part of the LLVM~\cite{lattner2004llvm},
which supports any language that can be compiled to LLVM intermediate representation. % a compiler frontend is implemented.
Currently, OSS-Fuzz~\cite{serebryany2017oss} supports C/C++, Rust, Go, Python, and Java/JVM code,
and other LLVM-supported languages.

Syntax transformation from fuzz targets to unit test templates differs for languages.
However, the general framework can be defined in a language-agnostic manner.
\unitsyn~\cite{he2024unitsyn} is a multi-lingual framework to collect unit test functions
based on tree-sitter,
which can be extended to syntax transformation.
% If further research finds a native compiler as a more desirable tool,
% an approach similar to ours can also be applied using
% \texttt{ast}\footnote{\url{https://docs.python.org/3/library/ast.html}}
% module for Python datasets.

We choose Rust~\cite{matsakis2014rust} to conduct our study to take advantage of its powerful build tool \texttt{cargo}\footnote{\url{https://doc.rust-lang.org/cargo/}}.
\texttt{Cargo-fuzz}\footnote{\url{https://github.com/rust-fuzz/cargo-fuzz}}
allows software developers to define their fuzz targets inside the repository,
making it easier for us to execute the fuzz targets and apply our data augmentation.
In principle, our method can be generalized to all libFuzzer-supported languages,
and their corresponding fuzz targets can be found in OSS-Fuzz~\cite{serebryany2017oss}.
To use \fuzzaug in other languages,
one could locate the fuzz targets in OSS-Fuzz.
The current limitation of \fuzzaug is that only languages supported by OSS-Fuzz can be used.

\begin{comment}
For LLVM-supported languages like C/C++, libFuzzer is natively supported, and reporter instrumentation is implemented via LLVM's analysis and transform passes.
% allowing syntax transformation using \texttt{libclang}. 
Similarly, Java Virtual Machine (JVM) languages can benefit from fuzzing through Jazzer~\cite{jazzer}.
Using \texttt{javaagent} for reporter instrumentation,
\fuzzaug is applicable to Java datasets and unit test generation in Java. 
Python is supported through Google's \texttt{atheris}~\cite{google/atheris} for fuzzing, 
with syntax transformation leveraging the \texttt{ast} module and decorators for instrumentation. 
By supporting LLVM, JVM, and Python, \fuzzaug provides broad compatibility, 
advancing data augmentation across widely used programming languages and contributing significantly to program testing by deep language models.
\end{comment}

\subsection*{Applying to Different Datasets}

We followed TeCo~\cite{nie2023TeCo} and \unitsyn~\cite{he2024unitsyn}
to construct our dataset on function-level code-test pairs.
% In our training setting, 
% each training example consists of a focal function and a unit test function concatenated by a newline symbol, 
% providing efficiency and strong performance. 
File-level pairing approach used in CAT-LM~\cite{rao2023catlm} offers additional benefits by providing more relevant context, 
which is particularly useful in less modular, tightly coupling, complex software systems. 
\fuzzaug is applicable to both function-level and file-level data
to accommodate various types of datasets effectively.
% To accommodate both function-level and file-level data, 
% a robust data augmentation method must be applicable to both levels. 
% \fuzzaug satisfies this requirement by supporting file-level pairing, 
LibFuzzer maintains separate fuzz targets in different files. 
After syntax transformation and fuzz data collection, 
\fuzzaug can insert augmented unit test functions into their original files and adopt CAT-LM's pairing strategy. 
This versatility enhances \fuzzaug's ability to augment and improve various types of unit test datasets effectively.
However, \fuzzaug requires the software repositories to compile successfully.

\subsection*{Evaluation on Real-World Projects}

In our experiments, we follow \unitsyn to assess the validity and completeness of generated unit test functions using HumanEval-X~\cite{zheng2023codegeex}.
We did not use real-world Rust projects due to a few challenges.
% The first is the data leakage issue discussed in \unitsyn.
First, as discussed in UniTSyn,
it is hard to eliminate data leakage when evaluating on open-source projects.
\citet{he2024unitsyn} conducted a detailed analysis of the data leakage issue,
and conclude that user their dataset construction method,
there will be no data leakage on HumanEval-X in the training process.
%  in training and evaluating on the same open-source projects,
% as discussed by \citet{he2024unitsyn}.

\begin{comment}
First, we try to eliminate data leakage.
The training dataset of the state-of-the-art code generation models is likely to contain the popular repositories on GitHub,
including the ones we used to build our dataset.
Therefore, evaluating the models on the repositories we collected from GitHub
might be unfair due to data leakage.
Although comparing with other state-of-the-art models is not our goal,
the effect of data leakage on evaluating \fuzzaug is unknown.
To minimize such effects,
we choose to use a hand-crafted benchmark,
and HumanEval-X is the only hand-crafted benchmark we found that contains Rust problems.
Although there might still be data contamination problem in the pre-training stage,
we believe the effect is limited, as we are letting the models to generate unit tests instead of the focal functions.
\end{comment}

Second, we want to minimize the negative impacts of incorrect project setup.
Generating unit tests in large open-source software (OSS) requires special setups for each project.
These setups for defect testing are hard to construct and require human domain knowledge~\cite{zhu2023reproducibility}.
Therefore, choosing to evaluate test generation on OSS introduces additional bias in the results,
which is another thing we want to eliminate.

\begin{comment}
Third, Rust programs are hard to compile in a complex codebase.
Different from other popular languages in code generation,
for example, Python and JavaScript with dynamic types and C/C++ with weak types,
Rust's type system is both static and strong,
making code generation in Rust hard to pass the type checking and therefore fail to compile.
In addition to static and strong typing,
Rust introduced linear logic~\cite{girard1987linear} (linear types~\cite{wadler1990linear})
into its type system to manage the lifetime and ownership of pointers,
adding additional difficulty in generating type-correct code in the complex codebase.
To this end, we found HumanEval-X to be a more suitable benchmark since
the problems are simple enough without any over-complex types,
yet are challenging enough that the focal functions contain multiple branches to evaluate code coverage.
As a result, we select HumanEval-X to evaluate \fuzzaug instead of using the OSS projects.
\end{comment}

Finally, a hand-crafted and expert-verified benchmark like HumanEval-X offers
an oracle implementation of the focal functions.
If we use real-world projects to evaluate LLM-based unit test generation and an assertion failed,
we have no directly way to distinguish whether the generated unit test is incorrect or there is an actual defect.
Previous work~\cite{pacheco2007randoop} in automated unit test generation
uses very simple assertions as oracles, such as \rscode{assert o.equals(o)},
aimed at finding bugs in codebases.
Our goal is to evaluate the completeness and correctness of the generated unit test functions,
so we need a benchmark that can provide the oracle implementation of the focal functions.
One interesting future work direction is to construct a ground-truth benchmark on selected real-world projects for neural test generation,
where all the bugs are known and the oracle implementation is available.
Examples in this direction include BugSwarm~\cite{tomassi2019bugswarm} and Magma~\cite{magma}.

%% file: src/appendix.tex
\section{Appendix}

\subsection{Training Details} \label{sec:training_details}

\begin{figure}[ht]
	\centering
	\includegraphics[width=\columnwidth]{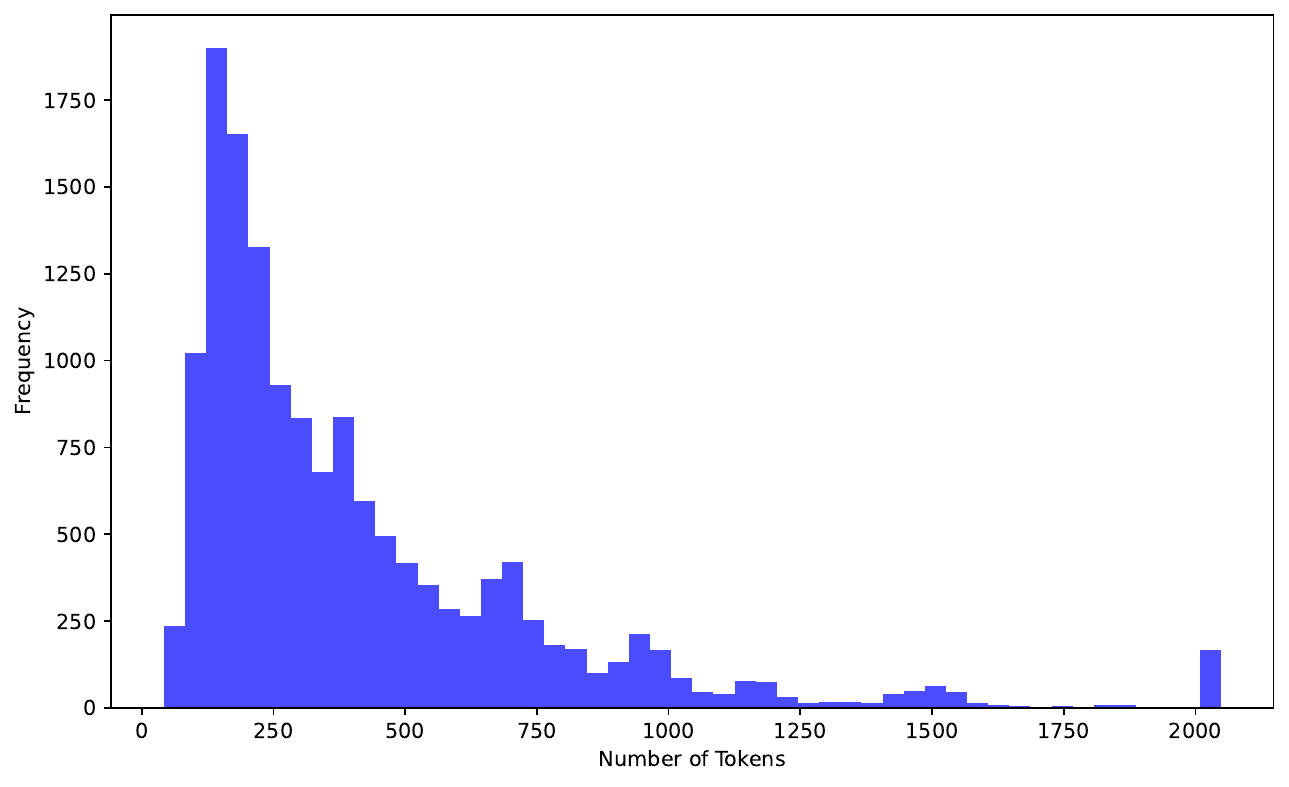}
	\caption{Token distribution of the dataset.}
	\label{fig:token_distribution}
\end{figure}

We follow the previous work~\cite{radford2019language,he2024unitsyn}
to use an autoregressive signal for continual training of the pre-trained base model.
We follow \unitsyn for the basic training configuration.
Specifically,
each training example is the concatenation of the focal function and the unit test function,
joined by a \texttt{\textbackslash n} new line symbol.
Since most of the training data is around 250 tokens (see \autoref{fig:token_distribution}),
we set the maximum sequence length to 512 for the tokenizer.
We use a batch size of 128, with gradient accumulation at every 32 steps.
We use a $5\mathrm{e}^{-5}$ learning rate for our training,
with cosine annealing learning rate decay for each batch~\cite{loshchilov2016sgdr}.
Following \citeauthor{kirkpatrick2017overcoming}, we use 0.05 weight decay
to make the trained model robust to catastrophic forgetting.
We apply LoRA~\cite{hu2022lora} to the model with the rank $r = 16$, $\alpha = 16$, and 0.05 dropout.
We train all the models, except StarCoder2, for 100 steps (approximately eight epochs) 
on four NVIDIA H100-80GB GPUs.
StarCoder2 is trained for 200 steps due to its slower convergence rate and poor performance.

\subsection{Testing in Practice} \label{sec:testing_format}

Unit testing is a software testing technique that focuses on assessing the correctness of basic software units~\cite{zhu1997unittest}.
In classical setups, unit tests contain three major stages: arrange, act, and assert~\cite{khorikov2020unit}.
The arrange stage sets up the input data in the correct format,
the act stage invokes the code under test,
and the assert stage checks the output of the code.
If passed, these unit tests can be used as regression tests to ensure the future correctness and security of the software~\cite{pacheco2007randoop}.
Unit tests in software repositories are usually structured as \emph{test functions},
each encapsulating the semantics of the aforementioned three components.
Unit test functions can be identified using language-specific hooks~\cite{he2024unitsyn}.

\paragraph{Unit Testing in Rust.}
Unit testing in Rust is no different from that in other programming languages.
Rust provides a built-in test framework that allows developers to specify unit test functions using the \rscode{#[test]} or \rscode{#[cfg(test)]} attribute.
The \texttt{rustc} compiler can automatically identify these test functions at compile time and includes them only in the test build.
Rust offers assertions through the \rscode{assert!} macro,
with variants such as \rscode{assert_eq!} and \rscode{assert_ne!} for checking equality and inequality, respectively.
These assertion macros are used to verify the expected behavior of the code when the tests are executed.
An example of a Rust unit test function is shown in \autoref{lst:rust:unit_test},
illustrating a simple arrangement on the first line,
followed by the action and assertion within the \rscode{assert_eq!} macro on the next line.

\paragraph{Fuzzing in Rust.}
The \texttt{cargo-fuzz} tool provides fuzzing functionality for Rust using LibFuzzer~\cite{serebryany2016libfuzzer}.
However, instead of being defined as a test function,
a fuzz target is specified using the \rscode{fuzz_target!} macro, which takes a closure function as an argument.
The closure function provides the appropriate testing semantics.
Unlike unit test functions,
where programmers hardcode test inputs during the arrange stage,
fuzz targets supply randomized input data of type \rscode{&[u8]} (a slice of 8-bit unsigned integers) to the closure function.
The closure function is then responsible for correctly parsing the input into the appropriate format for the arrange stage.
After that, the closure function follows the same semantics as a unit test function:
the act stage invokes the code under test, and the assert stage verifies its output.
As shown in the example in \autoref{lst:rust:fuzz_target},
the closure function performs the arrange stage on line 7.
This key design of fuzz targets enables syntax transformation to convert a fuzz target into a unit test function,
as described in \autoref{sec:transformation}.

\subsection{Additional Results}

\input{tabels/assert_additional.tex}
\input{tabels/func_addational.tex}

\subsection{Additional Figures}
\input{code/helpers.tex}
\begin{figure*}[t]
	\centering
	\includegraphics[width=0.7\textwidth]{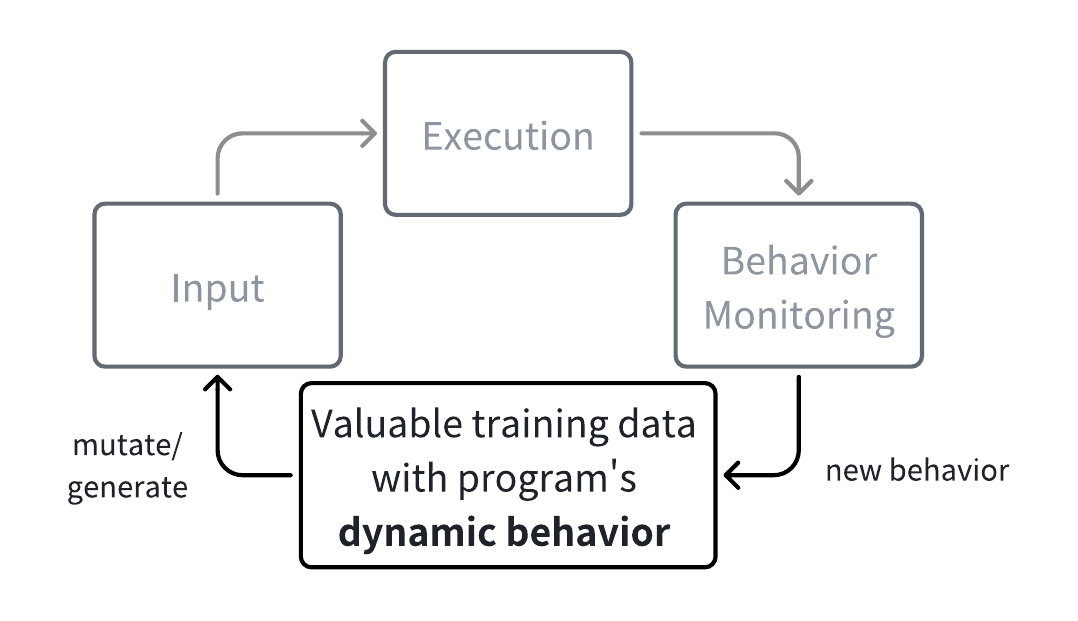}
	\caption{Fuzzing loop for dynamic program testing.
		This loop shows the process of the collection of randomized generated data for augmentation.}
	\label{fig:fuzzing}
\end{figure*}

\begin{figure*}[h]
    \centering
    \includegraphics[width=\textwidth]{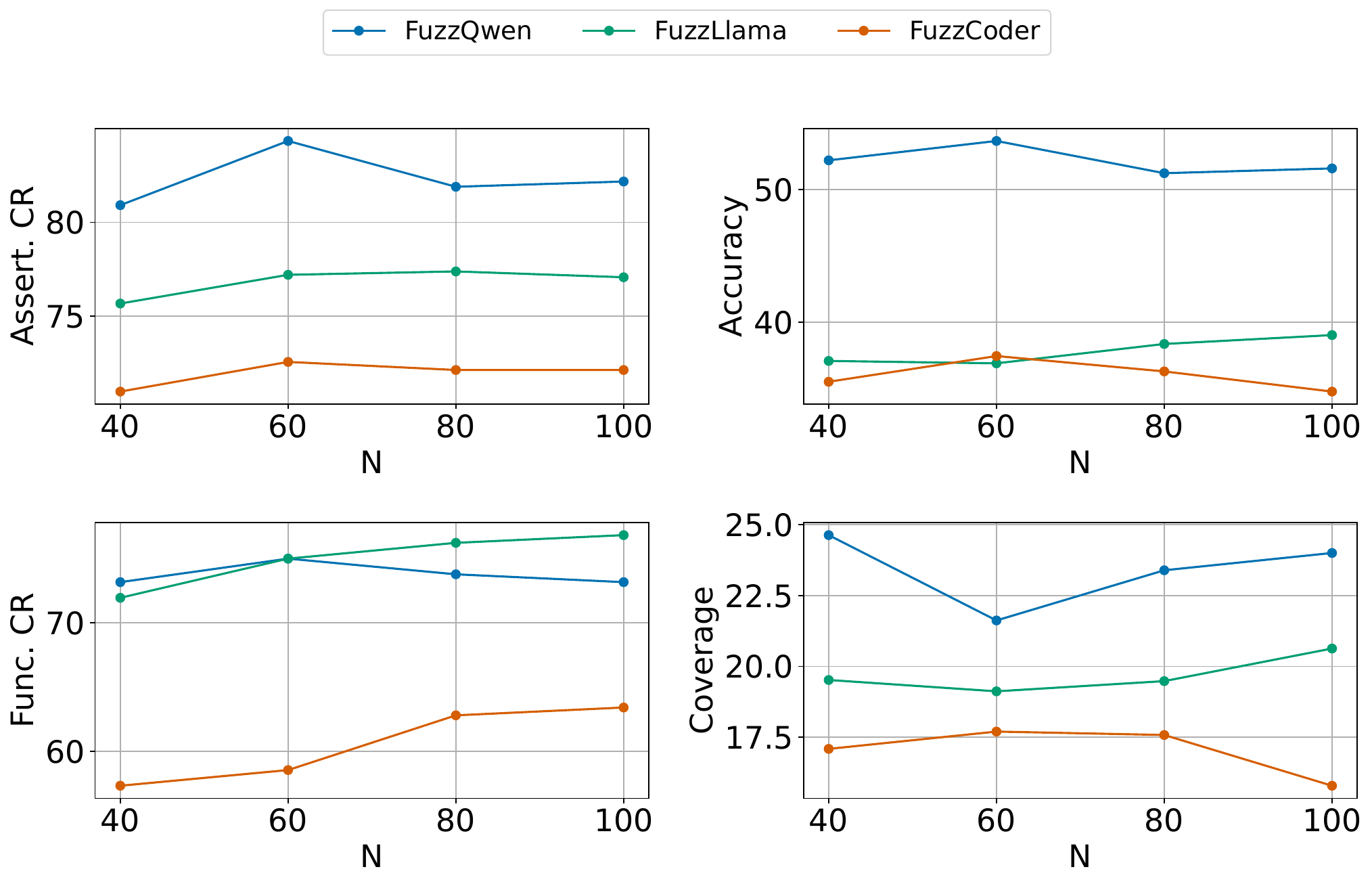}
    \caption{The impact of scaling the number of sampled fuzzing inputs on test generation performance.}
    \label{fig:scaling}
\end{figure*}

%% file: tabels/assert_additional.tex
\begin{table}[h]
	\centering
	\begin{tabular}{lccc}
		\toprule
		Model & Type & Assert. CR & Acc   \\
		\midrule
		GPT-4 & API  & 95.53      & 75.04 \\
		\bottomrule
	\end{tabular}
	\caption{
		Accuracy of tests generated by LLMs. The best results are highlighted in bold.
		Assert. CR: the compile rate of the individual assertions.
		Acc: accuracy of individual assertions.
	}
	\label{tab:assert_addational}
\end{table}

%% file: tabels/func_addational.tex
\begin{table}[h]
	\centering
	\begin{tabular}{lccc}
		\toprule
		Model & Type & Func. CR & Cov   \\
		\midrule
		GPT-4 & API  & 93.90    & 47.94 \\
		\bottomrule
	\end{tabular}
	\caption{Evaluations of usefulness of generated unit tests.
		Func. CR: the compile rate of generated unit test functions.
		Cov: the average branch coverage of generated unit test functions on the focal functions.
	}
	\label{tab:func_addational}
\end{table}

%% file: code/helpers.tex
\begin{algorithm*}[t]
\caption{Fuzzing as Data Augmentation}
\label{alg:helpers}
\begin{algorithmic}[1]
\Function{ReporterInstrumentation}{$fuzz\_target$}
    \State $AST \gets$ \textproc{Parse}($fuzz\_target$) % \Comment{Pointer to AST for mutation} % Using * to indicate pointer
    \State $entry \gets$ \textproc{GetBegin}($AST$) \Comment{Pointer to the entry point}% Consistently using * to indicate pointer usage
    \State $data \gets$ \textproc{GetParameters}($AST$)[0]
    \State $AST' \gets$ \Call{AddInstruction}{$AST$, $entry*$, \textproc{Report}($data$)} \Comment{Add reporter the entry of AST}
    \State $fuzz\_target' \gets$ \textproc{Dump}($AST'$)
    \State \Return $fuzz\_target'$
\EndFunction
\Statex
\Function{SyntaxTransformation}{$fuzz\_target$}
    \State $AST* \gets$ \textproc{Parse}($fuzz\_target$)
    \State $body \gets$ \textproc{ExtractBodyNode}($AST*$)
    \State $test\_header \gets ...$ \Comment{Language-specific header}
    \State $data\_template \gets ...$ \Comment{Declaring data variable}
    \State $test\_ending \gets ...$ \Comment{Closing this test definition}
    \State \Return $test\_header$ + $data\_template$ + $body$ + $test\_ending$
\EndFunction
\Statex
\Function{\fuzzaug}{$repo$, $N$, $L$, $timeout$}
    \\\Comment{$repo =$ repository to apply \fuzzaug}
    \\\Comment{$N =$ number of training examples to generate}
    \\\Comment{$L =$ maximum input length for collection}
    \\\Comment{$timeout =$ maximum allowed fuzzing time}
    \State $dataset_{\text{aug}} \gets$ []
    \ForAll{$t \in$ \textproc{GetFuzzTarget}($repo$)}
        \State $t' \gets$ \textproc{ReporterInstrumentation}($t$)
        \State $inputs \gets$ \textproc{Fuzz}($t'$, $timeout$) \Comment{Collect raw fuzzing inputs}
        \State $inputs' \gets$ \textproc{Filter}($\lambda x:$ \textproc{len}($x$) $< L$, $inputs$)
        \State $selected \gets$ \Call{Sample}{$N$, $inputs'$}
        \State $templates \gets$ \textproc{Take}($N$, \Call{SyntaxTransformation}{$t$})
        \State $dataset_{\text{aug}} \gets$ $dataset_{\text{aug}}$ + \Call{Instantiate}{$templates$, $selected$}
    \EndFor
    \State \Return $dataset_{\text{aug}}$
\EndFunction
\end{algorithmic}
\end{algorithm*}